\documentclass[12pt]{iopart}
\usepackage[dvips]{graphics, color}
\usepackage{dcolumn}
\usepackage{iopams,wrapfig}
\usepackage{xspace}
\usepackage{epsfig}

\def\be {\begin{equation}}
\def\ee {\end{equation}}

\def\bea {\begin{eqnarray}}
\def\eea {\end{eqnarray}}

\newcommand{\bef}{\begin{figure}}
\newcommand{\eef}{\end{figure}}

\begin{document}
\title[Electromagnetic probes]{Measuring initial temperature through photon 
to dilepton ratio in heavy ion collision.}

\author{Jajati K. Nayak, Jan-e Alam, Sourav Sarkar and Bikash Sinha}
\medskip
\address{ Variable Energy Cyclotron Centre, 1/AF Bidhan Nagar,
Kolkata 700 064, INDIA}
\ead{jajati-quark@veccal.ernet.in}

\begin{abstract}
Theoretical calculation of transverse momentum($p_T$) distribution of 
thermal photons and dileptons originating from ultra-relativistic 
heavy ion collisions suffer from several uncertainties since the 
evaluation of these spectra needs various inputs which are not yet known 
unambiguously. In the present work the ratio of the $p_T$ spectra of 
thermal photons to lepton pairs has been evaluated and it is shown that the 
ratio is insensitive to some of these parameters.  
\end{abstract}

\section{Introduction}
It is expected that the collision of heavy ions at ultra-relativistic 
energies would create a thermalised system of elementary particles like 
quarks and gluons. The interaction of these quarks and gluons produces real 
and virtual photons (dileptons). In principle, photons and dileptons 
can be used as effective tools to understand the initial state of the 
matter formed in heavy ion collision~\cite{larry,GK,weldon90,jpr}. 
However, in practice, difficulties arise firstly in the separation of 
thermal radiation from quark gluon plasma(QGP) and from those produced 
in initial hard collisions and from the decays of hadrons. 
Secondly, the evaluation of thermal photon and dilepton 
spectra need various inputs such as initial temperature($T_i$), 
thermalisation time($\tau_i$), equation of state(EOS), 
critical temperature($T_c$) for quark-hadron phase transition, 
freeze-out temperature($T_f$) etc., which are not known unambiguously.
In the present work we evaluate the ratio,
$R_{em}=(d^2 N_{\gamma}/d^2 p_T dy)_{y=0}/(d^2 N_{\gamma^{\star}}/
d^2 p_T dy)_{y=0}$, of thermal photons to dileptons for different 
initial energy densities and show that some of the uncertainties 
mentioned above get canceled in the ratio.
\par
In section 2 the invariant yield of thermal photons and lepton pairs 
have been discussed. The dynamics of the space time evolution have 
been discussed in section 3. Finally, section 4 is devoted to 
results and discussions. 

\section{Production of thermal photons and lepton pairs} 
The $p_T$ distribution of thermal photons from heavy ion collisions 
in a first order phase transition scenario can be written as
\begin{equation}
\frac{d^2N_\gamma}{d^2p_{T}dy}=\sum_{i=Q,M,H}{\int_{i}{\left(\frac{d^2R_\gamma}
{d^2p_{T}dy}\right)_id^4x}}
\label{eq1}
\end{equation}
where $Q, M, H$ represents QGP, mixed (coexisting phase of QGP 
and hadrons)and hadronic phases respectively. $(d^2R/d^2p_{T}dy)_i$ is 
the production rate of photon from the phase $i$ at a temperature $T$, 
which is convoluted with the expansion dynamics through space-time 
integration over $d^4x$. The complete calculation for emission rate of 
photons from QGP to O($\alpha \alpha_s$) as done in ref ~\cite{arnold} 
has been considered in the present work. The rate of photon 
production in the hadronic phase has been taken from 
~\cite{we1,we2,we3,turbide}. Similarly the invariant transverse 
momentum distribution of thermal dileptons is given by:
\begin{equation}
\frac{d^2N_{\gamma^\ast}}{d^2p_{T}dy}=\sum_{i=Q,M,H}{\int_{i}
{\left(\frac{d^2R_{\gamma^\ast}}{d^2p_{T}dydM^2}\right)_idM^2d^4x.}}
\label{eq3}
\end{equation}
The limits for integration over $M$ can be fixed from the experimental 
measurements. In the plasma phase the lepton pair production 
rates to $O(\alpha^2 \alpha_s)$ have been considered 
~\cite{altherr,thoma}. The decays of vector mesons 
$\rho \rightarrow e^{+}e^{-}, \omega \rightarrow e^{+}e^{-}$ and 
$\phi \rightarrow e^{+}e^{-}$ have been considered for the dilepton 
production in the hadronic phase. See ~\cite{we5} for details.
\section{Space-time evolution}
The space time evolution of the system
has been studied using ideal relativistic hydrodynamics
assuming longitudinal boost invariance and cylindrical 
symmetry ~\cite{von,bjorken}. The initial temperature($T_i$)
and thermalisation time ($\tau_i$) are constrained
by the following equation for an isentropic expansion:
\be
T_i^{3}\tau_i \approx \frac{2\pi^4}{45\zeta(3)\pi R_A^2 4a_{k}}\frac{dN}{dy}.
\label{eq6}
\ee
where, $dN/dy$= hadron multiplicity, $R_A$ is the radius of the system,
$\zeta(3)$ is the Riemann zeta function and $a_k=\pi^2 g_{eff}/90$, 
$g_{eff}$ being the degeneracy of the initial phase. In the present
 work we assume $T_c = 192$ MeV. We use the Bag model EOS
for the QGP phase. For the EOS of the hadronic matter, 
all the resonances with mass $\leq 2.5$ GeV have been considered
~\cite{bm}. Electromagnetic spectra have also been evaluated with 
lattice EOS ~\cite{MILC} and results are compared with those obtained 
from phenomenological EOS (i.e, bag model for QGP and 
hadronic resonance gas for low temperature hadron phase).
\begin{figure}
  \begin{center}
    \rotatebox{0}{\resizebox{15.cm}{!}{\includegraphics{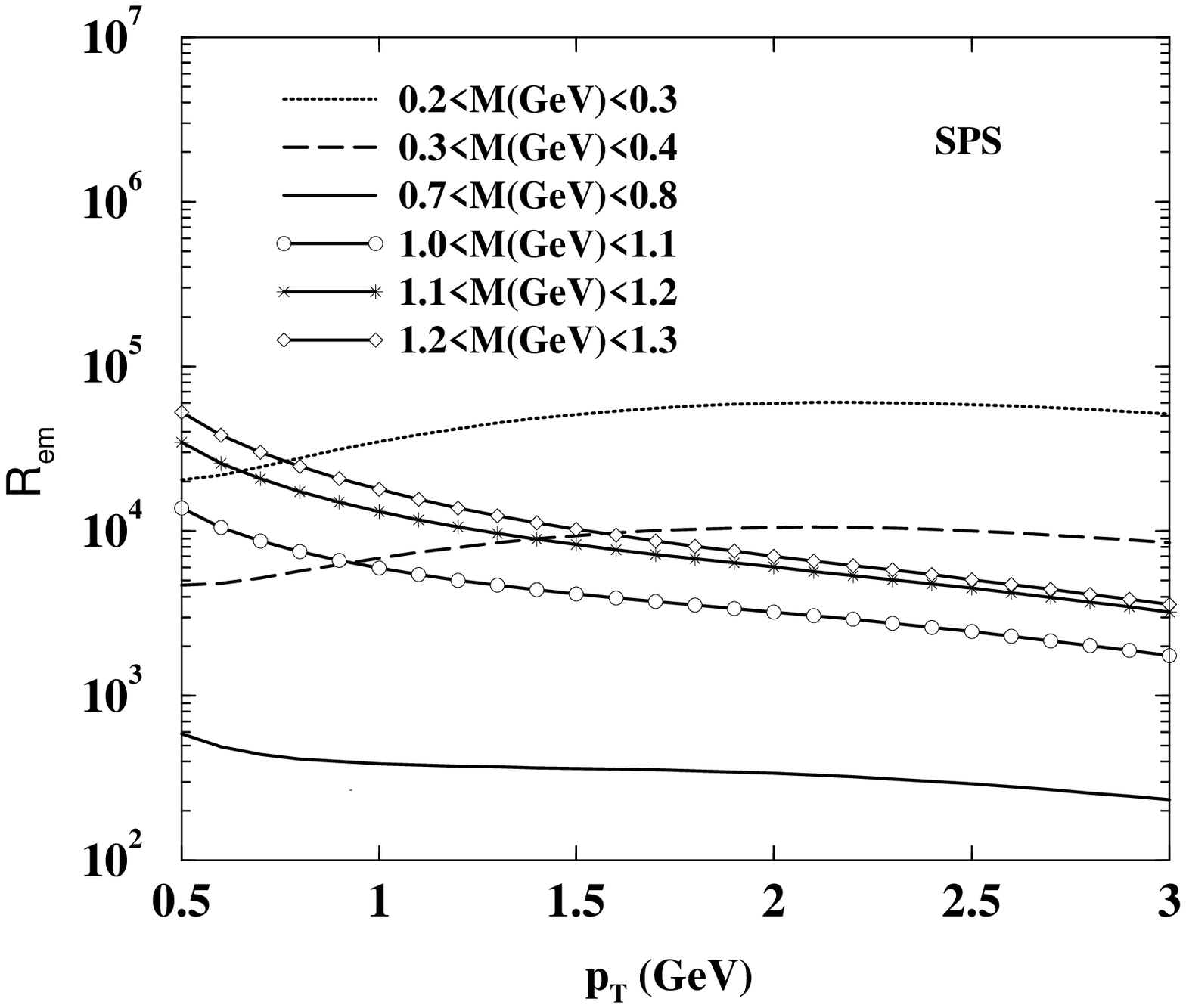}\includegraphics{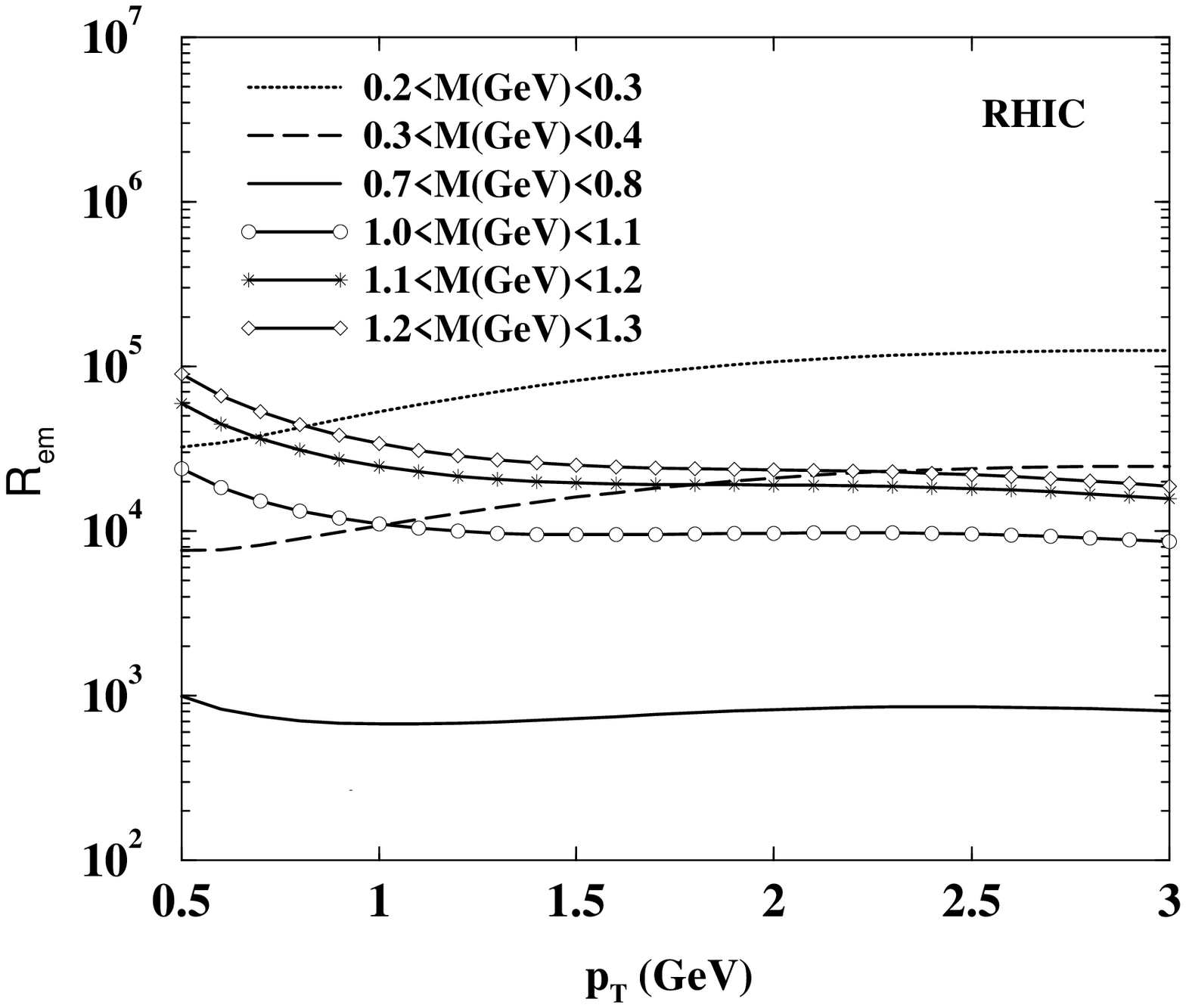}}}
    \vspace{-0.1cm} 
   \begin{tabular*}{15.7cm}{ @{\extracolsep{6.5cm}} l l @{\extracolsep{\fill}}}
     \hspace{0.5cm} \textbf{(a)} & \textbf{(b)}
   \end{tabular*} 
   \caption{The ratio, $R_{em}$ as function of $p_T$ is shown for 
different mass windows (a) for SPS energy
(b)RHIC energy}
  \end{center}
\end{figure}

\begin{figure}
  \begin{center}
    \rotatebox{0}{\resizebox{15.cm}{!}{\includegraphics{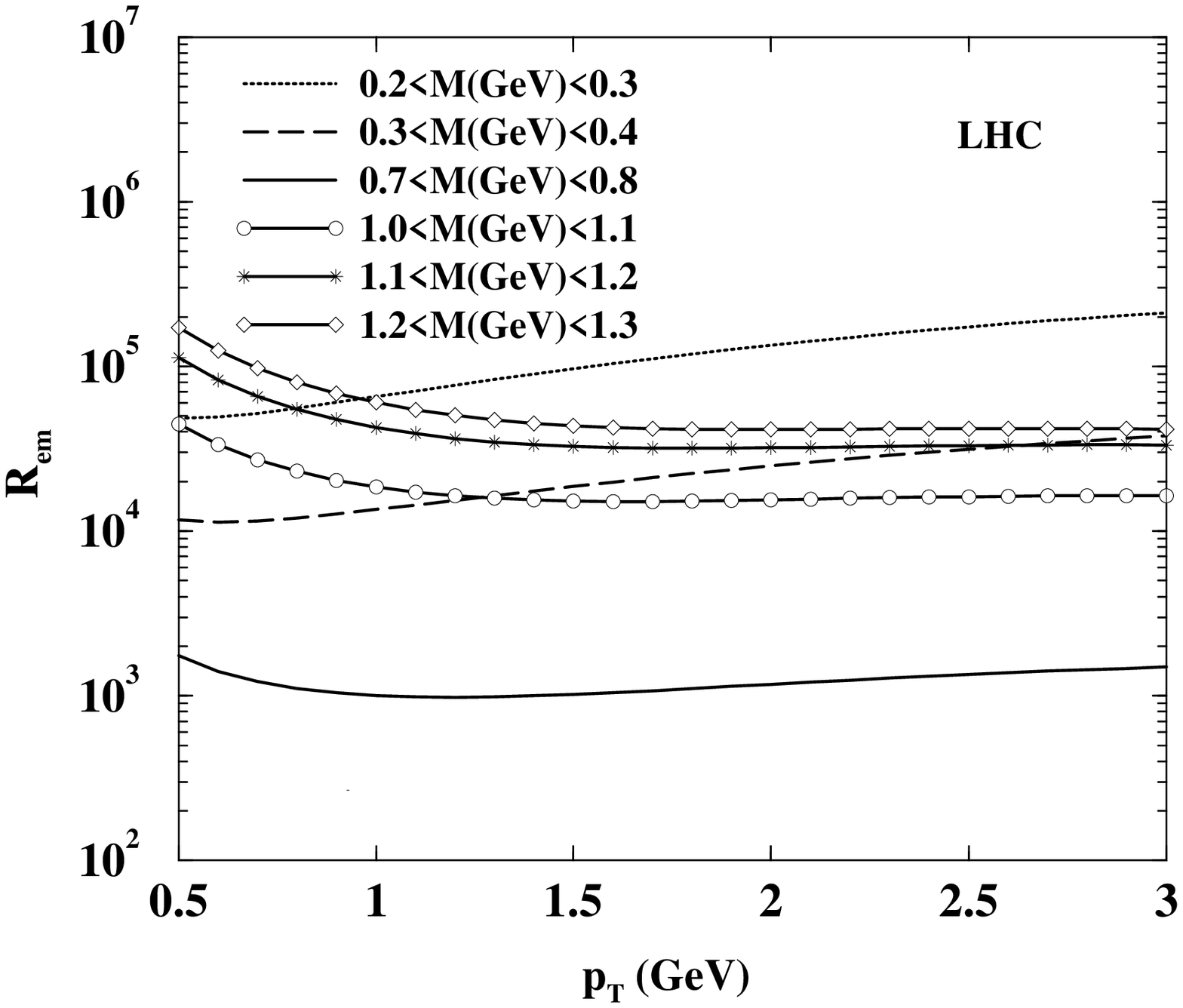}\includegraphics{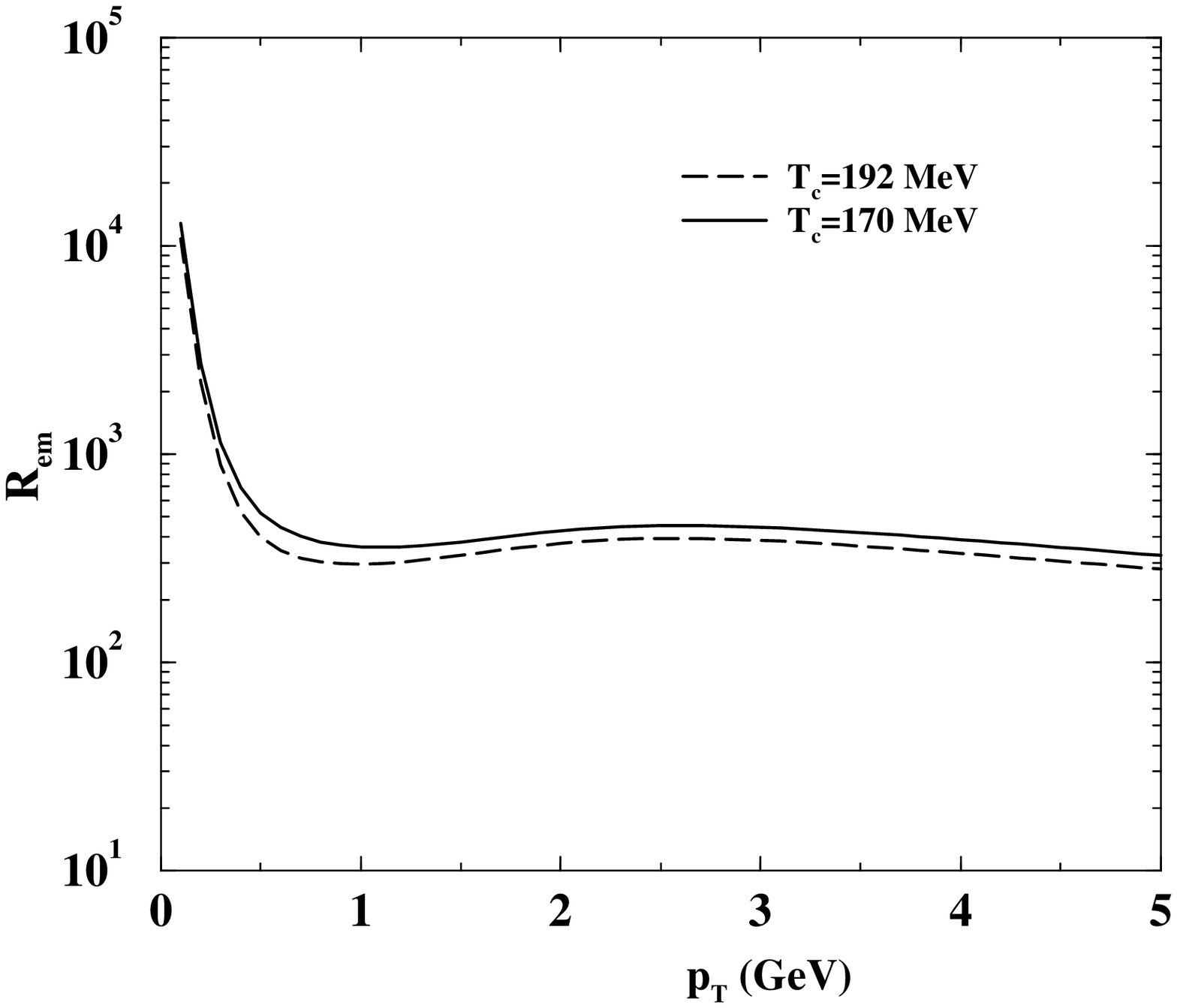}}}
    \vspace{-0.1cm} 
   \begin{tabular*}{15.7cm}{ @{\extracolsep{6.5cm}} l l @{\extracolsep{\fill}}}
     \hspace{0.5cm} \textbf{(a)} & \textbf{(b)}
   \end{tabular*} 
   \caption{(a) Ratio at LHC energy 
(b) The sensitivity of $R_{em}$ to $T-c$ 
when dilepton spectra is integrated over $M=2m_{\pi}$ to $m_{\phi}$}

  \end{center}
\end{figure}
\section{Results and Discussions}
The input parameters considered for SPS, RHIC and LHC energies 
have been shown in table-I. The invariant mass spectra of dileptons and 
transverse momentum spectra of photons measured by CERES~\cite{ceres} and WA98 
~\cite{wa98} at SPS energies have been reproduced~\cite{alam} with the 
initial conditions mentioned in table-I. The photon spectra measured by PHENIX 
collaboration~\cite{phenix} has also been explained with these input 
parameters ~\cite{alam} as shown in table-I for RHIC. 
\begin{table}
\caption{The values of various parameters - $\tau_i$, $T_i$, $T_f$ 
and hadronic multiplicity $dN/dy$  - used
in the present calculations.}
\begin{tabular}{lcccr}
Accelerator &$\frac{dN}{dy}$&$\tau_i(fm)$&$T_i$(GeV) &$T_f$ (MeV)\\
SPS&700&1&0.2&120\\
RHIC&1100&0.2&0.4&120\\
LHC&2100&0.08&0.70&120\\
LHC&4000&0.08&0.85&120\\
\end{tabular}
\end{table}
The ratio, $R_{em}$, of the transverse momentum spectra of thermal 
photons to dileptons has been evaluated for SPS, RHIC and LHC energies. 
It is observed that the ratio reaches a plateau beyond 
$p_T$= 1 GeV if the dilepton spectra is integrated from 
$M$=$2m_{\pi}$ to $m_{\phi}$ ~\cite{we5}. 
The ratio for different  mass windows are shown in figures 1a, 
1b and 2a for SPS, RHIC and LHC energies respectively. Although the 
individual spectra shows a strong dependence on the parameters 
mentioned above, the effect on the ratio is small, e.g. if we change 
$T_c$ from 192 to 170 MeV then the invariant yield of photons at 
RHIC energy changes by 8.5\% and dilepton spectra by 13\% at $p_T= 2 GeV$.
In contrast, $R_{em}$ changes by 5\%. $R_{em}$ depends on $T_c$ 
very weakly as evident from the results shown in figure 2b for RHIC. 
The effect of flow and medium effects on hadrons ~\cite{BR} on the 
ratio is negligibly small.  The transverse mass distribution of 
the dileptons for various invariant mass windows measured by 
NA60~\cite{na60} and PHENIX~\cite{phenix} collaborations 
could also be reproduced with the initial 
conditions shown in table-I. The details will be published elsewhere. 


\end{document}